\documentclass[preprint,showpacs,preprintnumbers,amsmath,amssymb]{revtex4}

\usepackage{graphicx}
\usepackage{dcolumn}
\usepackage{bm}

\begin{document}

\preprint{APS/123-QED}

\title{Superfluid-Mott Insulator Transition of Spin-1 Bosons in an Optical Lattice}

\author{Shunji Tsuchiya}
 \email{tuchiya@kh.phys.waseda.ac.jp}
\author{Susumu Kurihara}%
\affiliation{%
Department of Physics, Waseda University 3-4-1 Okubo, Tokyo 169-8555, Japan
}%

\author{Takashi Kimura}
\email{kimura@cms.phys.s.u-tokyo.ac.jp}
\affiliation{Department of Physics, University of Tokyo, 
Tokyo 113-0033, Japan, 
Faculty of Frontier Sciences, University of Tokyo, 
Tokyo 113-0033, Japan, and Advanced Research Institute for Science and Engineering, Waseda University, 3-4-1 Okubo, Tokyo 169-8555, Japan
}%

\date{\today}

\begin{abstract}
We have studied superfluid-Mott insulating transition of 
spin-1 bosons interacting antiferromagnetically in an optical lattice.
We have obtained the zero-temperature phase diagram 
by a mean-field approximation and have found that 
the superfluid phase is to be a polar state as 
a usual trapped spin-1 Bose gas. 
More interestingly, we have found that 
the Mott-insulating phase is strongly stabilized  
only when the number of atoms per site is even.

\end{abstract}

\maketitle


Superfluid-Insulator (S-I) transition has attracted 
attention, and has been extensively studied in the context of $^4\rm{He}$ 
absorbed in the porous media \cite{Reppy, Fisher}, granular superconductors 
\cite{Goldman} and Josephson-junction arrays \cite{Fazio}. 
Recently, 
Greiner {\it et al.} \cite{Greiner} has observed an S-I transition 
of $^{87}\rm Rb$ atoms trapped in a three-dimensional optical 
lattice potential by changing the potential depth 
when the number of atoms par site is an integer. 
This method is the most ideal way to study the S-I transition. 
There are no lattice imperfections, 
and we can easily change the potential depth 
in order to study both superfluid 
and insulator phases in a single system. 

On the other hand, recent advances of experimental techniques 
in an optical trap \cite{Stenger,Chapman} 
have achieved condensation of spinor bosons. 
Recent theoretical studies predict a variety of novel phenomena 
of spinor condensates such as fragmented condensation \cite{HoYip}, 
skyrmion excitations \cite{Ho,Tuchiya,Khawaja,Mizushima} 
and propagation of spin waves \cite{Ho,Ohmi}.  

Here, a natural question is what is expected 
if the spinor bosons are trapped in an optical lattice. 
In fact, several unique properties of spinor Bose atoms 
in an optical lattice have been suggested 
by Demler and Zhou \cite{Demler}. 
They have proposed some possible phases 
including the superfluid and insulating phases \cite{Demler},  
however, no microscopic calculation has been 
given for their boundaries nor stabilities. 
In this Letter, we study the S-I transition of spin-1 bosons 
with antiferromagnetic interaction in an optical lattice at 
zero temperature when the number of atoms per site is an integer. 
Using a mean-field approximation \cite{Stoof,Sheshadri}, 
we show the zero-temperature phase diagram 
where the superfluid phase is a polar state as in 
the case of spinor bosons trapped in a usual harmonic trap. 
More interestingly, we have found that 
the Mott-insulating phase is strongly stabilized  
only when the number of atoms per site is even. 


Bosons with hyperfine spin $F=1$, 
which include alkali atoms 
with nuclear spin $I=3/2$ such as $^{23}\rm{Na}$, 
$ ^{39}\rm{K}$ and $^{87}\rm{Rb}$, 
are represented by the Bose-Hubbard model\cite{Jaksch} 
in an optical lattice as 

\begin{eqnarray}
H=&&-t\sum_{<i,j>,\alpha}(a_{i \alpha}^\dagger a_{j \alpha}^{}
 + a_{j \alpha}^\dagger a_{i \alpha}^{}) -\mu\sum_{i,\alpha} a_{i \alpha}^\dagger a_{i \alpha}^{} \nonumber\\
&&+\frac{1}{2}U_0\sum_{i,\alpha,\beta} a_{i,\alpha,\beta}^\dagger a_{i \beta}^\dagger
a_{i \beta}^{}a_{i \alpha}^{}
+\frac{1}{2} U_2 \sum_{i,\alpha,\beta,\gamma} a_{i \alpha}^\dagger a_{i \gamma}^\dagger
{\mathbf F}_{\alpha \beta} \cdot {\mathbf F}_{\gamma \delta} a_{i \delta}^{}
 a_{i \beta}^{},\label{H}
\end{eqnarray}
where $a_{i \alpha}$ is the annihilation 
operator for an atom with hyperfine spin $\alpha$ ($=0,\pm1$) at site $i$ 
and $\mu$ is the chemical potential. 
$t=-\int d{\mathbf r} w_i^\ast({\mathbf r}) \left( -\hbar^2\nabla^2/2M
+V_0({\mathbf r})\right)w_j({\mathbf r})$ is the hopping matrix element between adjacent sites $i$ and $j$, where $w_i({\mathbf r})$ is a 
Wannier function localized on the $i$th lattice site, $M$ is the 
atomic mass and $V_0(\mathbf{r})$ 
is an periodic potential which characterizes an optical lattice. 
$U_0$ ($U_2$) is the on-site spin-independent (spin-dependent) 
inter-atom interaction. 
$U_F$ ($F=0,2$) is defined by $U_F=c_F 
\int d{\mathbf r}|w_i({\bf r})|^4$, where 
$c_0=(g_0+2g_2)/3$, $c_2=(g_2-g_0)/3$, $g_F=4\pi\hbar^2a_F/M$, and $a_F$
is an $s$-wave scattering length for two colliding atoms 
with total spin $F$. 
We assume an antiferromagnetic interaction $U_2>0$ 
($a_2>a_0$). This is the case for $^{23}\rm{Na}$ atoms.


In a Mott insulating phase with large inter-atom interaction 
($U_0$,$U_2\gg t$), we obtain an effective Hamiltonian within 
the second-order perturbation for the hopping parameter 
$t$ as 
$H_{\mathrm eff}=-J_1\sum_{<i,j>}{\mathbf{S}_i \cdot \mathbf{S}_j}
-J_2\sum_{<i,j>}\left(\mathbf{S}_i \cdot \mathbf{S}_j\right)^2=
-\frac{J_2}{4}\sum_{<i,j>}[(\mathbf{S}_i+\mathbf{S}_j)^2-4][(\mathbf{S}_i+ \mathbf{S}_j)^2-4+2/\alpha]$,
where
$J_1=2t^2/(U_0+U_2)$, 
$J_2=\frac{2t^2}{3}\left(\frac{1}{U_0+U_2}+\frac{2}{U_0-2U_2}\right)$, and $\alpha\equiv J_2/J_1=U_0/(U_0-2U_2)$.
If we consider the case with only two-sites, 
we obtain the spin-singlet (highest-spin) 
ground state if $U_2>(<)0$. 
In addition, it is known that at least in one-dimension, 
the ground state of this effective Hamiltonian 
is a dimerized state 
with a finite spin excitation gap if $U_2>0$,  
while the ground state is ferromagnetic state if $U_2<0$   
\cite{Schollwock}. 
The dimerized ground state with a positive $U_2$ suggests 
the polar state in the superfluid phase as we will see below. 


To study the S-I transition, we use a mean-field approximation 
in Refs. \cite{Stoof} and \cite{Sheshadri}.  
We start from $t=0$ case of Eq. \ref{H}, 
where the Hamiltonian is reduced to 
a diagonal matrix with respect to sites. 
Omitting the site index, the single-site Hamiltonian is 

\begin{eqnarray}
H_0=-\mu\hat{n} + \frac{1}{2}U_0\hat{n}(\hat{n}-1)
+\frac{1}{2}U_2({\hat{\mathbf S}}^2-2\hat{n}),
\end{eqnarray}
where $\hat{{\bf S}}=a_{\alpha}^\dagger {\mathbf F}_{\alpha \beta} a_{\beta}$ obeys a usual 
angular momentum commutation relation 
$[\hat{S}_i,\hat{S}_j]=i\epsilon_{ijk}\hat{S}_k$ 
and $\hat{n}=\sum_\alpha a_{\alpha}^\dagger a_{\alpha}$. 
$\hat{\mathbf{S}}^2$, $\hat{S_{z}}$, and $\hat{n}$ commute with each other. 
Therefore, the eigenstates of the above Hamiltonian 
are $|S,m;n\rangle(-S\le m \le S)$, where 
$\hat{\mathbf{S}}^2|S,m;n\rangle=S(S+1)|S,m;n\rangle$, 
$\hat{S_{z}}|S,m;n\rangle=m|S,m;n\rangle$, and 
$\hat{n}|S,m;n\rangle=n|S,m;n\rangle$. 
The energy of the eigenstate is $E^{(0)}(S,n)=-\mu n+\frac{1}{2}U_0n(n-1)
+\frac{1}{2}U_2[S(S+1)-2n]$.
Since the orbital wave function is symmetric,
the spin wave function has to be symmetric.
As a result, $S=0,2,4,...,n$, when the number of atoms $n$ is even and 
$S=1,3,5,...,n$, when $n$ is odd \cite{Bigelow}. 
The state with $m=S$ 
is $|S,S;n\rangle\propto (a_1^\dagger)^S 
(\Theta ^\dagger)^{(n-S)/2}|vac\rangle$
\cite{HoYip}, 
where $\Theta^\dagger\equiv {{a_0}^{\dagger}} ^2-2 {a_1}^{\dagger}{a_{-1}}^{\dagger}$ creates a spin-singlet pair. 
We obtain other states 
with lower magnetic quantum numbers
by operating 
$S^{-}=(S^{+})^\dagger$ to $|S,S;n\rangle$, 
where $S^{+}=S_{x}+ iS_{y}=\sqrt{2}({a_1}^{\dagger}a_0+{a_0}^{\dagger}a_{-1})$. 
Since we assume an antiferromagnetic interaction, 
the ground state is $|0,0;n\rangle$ 
with $n/2$ singlet pairs if $n$ is even, whereas 
the ground state is $|1,m;n\rangle(m=0,\pm1)$ 
if $n$ is odd. 
Comparing $E_{n}^{(0)}$, $E_{n+1}^{(0)}$ and $E_{n+2}^{(0)}$, 
we obtain the relation between the number of the atoms 
per site in the ground state 
and the chemical potential (Fig.1). Note that if $U_0<2U_2$, 
the atom number per site is even all over the phase diagram. 


Let us consider the case of finite $t$ to study the 
superfluid transition. 
Here, we introduce the superfluid order parameter  
$\psi_\alpha=\langle a_{i \alpha}\rangle=\sqrt{n_0}\zeta_\alpha$, 
where $n_0$ is the superfluid density and 
$\zeta_\alpha$ is a normalized spinor 
$\zeta_\alpha^\ast  \zeta_\alpha=1$. 
The hopping term is decoupled as 
$a_{i \alpha}^\dagger a_{j \alpha}\sim(\psi_\alpha a_{i \alpha}^\dagger+\psi_\alpha^\ast a_{j \alpha})-\psi_\alpha^\ast\psi_\alpha$.
As a result, the Hamiltonian is represented by a 
site-independent effective Hamiltonian multiplied 
by the total number of sites. The effective 
mean-field Hamiltonian \cite{Stoof,Sheshadri} is 
\begin{eqnarray}
H_\mathrm{mf}=&&H_0+zt\sum_{\alpha}\psi_\alpha^\ast \psi_\alpha+V,\nonumber\\ 
V=&&-zt\sum_{\alpha}(\psi_\alpha^\ast a_\alpha+\psi_\alpha a_\alpha^\dagger),
\end{eqnarray} 
where $z$ is the number of the nearest-neighbor sites.
$V$ corresponds to the transfer between bosons 
localized on a particular site and the superfluid.
We assume $\psi_\alpha$ and $t$ 
are small and include $V$ by a perturbation theory. 

We first consider the case when the number of atoms in a site is even. 
Calculating the second-order correction, 
we obtain the ground state energy as 
\begin{eqnarray}
&&\bar{E}_n(\psi)=\bar{E}^{(0)}(0,n)+A\left(n,\bar{U}_0,\bar{U}_2,\bar{\mu}\right)(\vec{\psi}^\dagger\cdot \vec{\psi}),\nonumber\\
&&A\left(n,\bar{U_0},\bar{U_2},\bar{\mu}\right)=\bigg[1+\frac{1}{3}\Big(\frac{n+3}{\bar{\mu}-\bar{U}_0 n}\nonumber\\
&&+\frac{n}{-\bar{\mu}+\bar{U}_0(n-1)-2\bar{U}_2}\Big)\bigg],
\end{eqnarray}
where $\bar{E}\equiv E/zt$, $\bar{\mu}\equiv \mu/zt$, $\bar{U}_F\equiv U_F/zt$ and  $\vec{\psi}\equiv\left(\psi_1,\psi_0,\psi_{-1}\right)$. 
Since the order parameter is determined 
to minimize the ground state energy, 
the ground state is the insulating (superfluid) phase
with zero (finite) $\vec{\psi}$ if $A>(<)0$.  
By the condition $A=0$, we obtain the upper ($\mu_+$) 
and lower ($\mu_-$) phase boundaries as 
\begin{eqnarray}
\bar\mu_\pm=&&-\bar{U}_2+\frac{1}{2}[(2n-1)\bar{U}_0-1]\nonumber\\
&&\pm\frac{1}{6}\Big\{9\bar{U}_0^2+6\left(6\bar{U}_2-2n-3\right)\bar{U}_0\nonumber\\
&&+[36\bar{U}_{2}^2-12(2n+3)\bar{U}_{2}+9]\Big\}^{1/2}.
\end{eqnarray}
By equating $\bar{\mu}_+$ and $\bar{\mu}_-$, 
we find the minimum of $\bar{U}_0$, denoting $\bar{U}_0^c$.   
The result is 
\begin{eqnarray}
\bar{U}_0^c=-\frac{1}{3}[6\bar{U}_2-(2n+3)]+\frac{2}{3}\sqrt{n^2+3n}. \label{U0C}
\end{eqnarray}

The results obtained by the second order perturbation with $V$ 
cannot determine the symmetry of the superfluid order parameter.
In order to determine this,  we have to calculate 
the fourth-order perturbation energy 
$E^{(4)}=\sum_{n,p,q\neq i}\langle i|V|n\rangle 
\frac{\langle n|V|p\rangle}{E_i^{(0)}-E_n^{(0)}}\frac{\langle p|V|q\rangle}{E_i^{(0)}-E_p^{(0)}}\frac{\langle q|V|i\rangle}{E_i^{(0)}-E_q^{(0)}}
-E^{(2)}\sum_{n} \frac{|\langle i|V|n\rangle|}{(E_i^{(0)}-E_n^{(0)})^2}$\cite{Messiah}.
A long but straightforward calculation gives 
 \begin{eqnarray}
\bar{E}_{n}^{(4)}=&&B\left(n,\bar{U}_0,\bar{U}_2,\bar{\mu}\right)n_0^2|\zeta_0^2-2\zeta_1\zeta_{-1}|^2\nonumber\\
&&+C\left(n,\bar{U}_0,\bar{U}_2,\bar{\mu}\right)(\vec{\psi}^\dagger\cdot \vec{\psi})^2,
\label{fourth}
\end{eqnarray}
with
\begin{eqnarray}
B\left(n,\bar{U}_0,\bar{U}_2,\bar{\mu}\right)=&&-\frac{1}{9}\bigg[\frac{n(n+1)}{\Delta \bar{E}^{(0)}(1,n-1)^2 \Delta \bar{E}^{(0)}(0,n-2)}\nonumber\\
&&+\frac{(n+2)(n+3)}{\Delta \bar{E}^{(0)}(1,n+1)^2 \Delta \bar{E}^{(0)}(0,n+2)}\bigg]\nonumber\\
&&+\frac{2}{45}\bigg[\frac{n(n-2)}{\Delta \bar{E}^{(0)}(1,n-1)^2 \Delta \bar{E}^{(0)}(2,n-2)}\nonumber\\
&&+\frac{(n+3)(n+5)}{\Delta \bar{E}^{(0)}(1,n+1)^2 \Delta \bar{E}^{(0)}(2,n+2)}\bigg]\nonumber\\
&&-\frac{n(n+3)}{15}\bigg[\frac{1}{\Delta \bar{E}^{(0)}(1,n+1)}\nonumber\\
&&+\frac{1}{\Delta \bar{E}^{(0)}(1,n-1)}\bigg]^2\frac{1}{\Delta \bar{E}^{(0)}(2,n)},\nonumber\\
C\left(n,\bar{U}_0,\bar{U}_2,\bar{\mu}\right)=&&-\frac{2}{15}\bigg[\frac{n(n-2)}{\Delta \bar{E}^{(0)}(1,n-1)^2 \Delta \bar{E}^{(0)}(2,n-2)}\nonumber\\
&&+\frac{(n+3)(n+5)}{\Delta \bar{E}^{(0)}(1,n+1)^2 \Delta \bar{E}^{(0)}(2,n+2)}\bigg]\nonumber\\
&&+\frac{n(n+3)}{45}\bigg[\frac{1}{\Delta \bar{E}^{(0)}(1,n-1)}\nonumber\\
&&+\frac{1}{\Delta \bar{E}^{(0)}(1,n+1)}\bigg]^2\frac{1}{\Delta \bar{E}^{(0)}(2,n)}\nonumber\\
&&+\frac{1}{9}\bigg[\frac{n}{\Delta \bar{E}^{(0)}(1,n-1)}+\frac{n+3}{\Delta \bar{E}^{(0)}(1,n+1)}\bigg]\nonumber\\
&&\bigg[\frac{n}{\Delta \bar{E}^{(0)}(1,n-1)^2}+\frac{n+3}{\Delta \bar{E}^{(0)}(1,n+1)^2}\bigg],
\end{eqnarray}
where $\Delta E^{(0)}(S,l)\equiv E^{(0)}(S,l)-E^{(0)}(0,n)$. 
The first term on the right-hand side of Eq. \ref{fourth} 
lifts the degeneracy of the superfluid order parameters in spin space. 
Because $B\left(n,\bar{U}_0,\bar{U}_2,\bar{\mu}\right)$ 
is negative with even $n$, 
the superfluid phase is a polar (spin-0) state 
$\vec{\zeta}=\left(0,1,0\right)$ as in 
the spin-1 Bose condensation in a usual harmonic trap \cite{Ho}. 
The fourth-order calculation is explained physically.  
The second intermediate states are $|0,0;n\pm2\rangle$, 
$|2,m;n\pm2\rangle$, and $|2,m;n\rangle$.  
The processes passing through 
$|0,0;n\pm2\rangle$ ($|2,m;n\pm2\rangle$) 
makes negative (positive) contributions 
to $B\left(n,\bar{U}_0,\bar{U}_2,\bar{\mu}\right)$ 
which favors (disfavors) the polar state.  
This is physically natural since 
$|0,0;n\pm2\rangle$ ($|2,m;n\pm2\rangle$) 
contains a spin-singlet (triplet) pair. 
Other processes passing through $|2,m;n\rangle$ make nontrivial
negative contributions 
to $B\left(n,\bar{U}_0,\bar{U}_2,\bar{\mu}\right)$. 
Adding all these contributions,  
we obtain 
a negative $B\left(n,\bar{U}_0,\bar{U}_2,\bar{\mu}\right)$, 
therefore the superfluid is the polar state. 


Now consider the case where the number of atoms per site is odd.
Since the non-perturbative ground state has 
degenerated states with $m=0$ and $\pm1$, 
we have to solve the secular equation 
$\Big|\langle1,m;n|V\frac{1}{E^{(0)}(1,n)-H_0}V|1,m';n\rangle\nonumber\\
-E_n^{(2)}\delta_{m m'}\Big|=0$
to lift the degeneracy.
The energy eigenvalues are obtained as
\begin{eqnarray}
\bar{E}_n^{(2)}=&&[-3(\beta+\delta)
-(\alpha+\gamma)+7(\beta+\delta)]
(\vec{\psi}^\dagger\cdot \vec{\psi}),\nonumber\\
&&\pm\Big\{[(\alpha-\gamma)-5(\beta-\delta)]^2 (\vec{\psi}^\dagger\cdot \vec{\psi})^2\nonumber\\
&&+4(3\beta+\gamma-2\delta)(\alpha-2\beta+3\delta)n_0^2|\zeta_0^2-2\zeta_1\zeta_{-1}|^2\Big\}^{1/2},\label{odd}
\end{eqnarray}
where $\alpha$,$\beta$,$\gamma$, and $\delta$ are given by 
$\alpha=\frac{n+2}{3}\frac{1}{\Delta \bar{E}^{(0)}(0,n-1)}$,
$\beta=\frac{n-1}{15}\frac{1}{\Delta \bar{E}^{(0)}(2,n-1)}$,
$\gamma=\frac{n+1}{3}\frac{1}{\Delta \bar{E}^{(0)}(0,n+1)}$, and
$\delta=\frac{n+4}{15}\frac{1}{\Delta \bar{E}^{(0)}(2,n+1)}$ 
respectively and $\Delta E^{(0)}(S,l)\equiv E^{(0)}(S,l)-E^{(0)}(1,n)$. 
The ground state energy corresponds to the lower sign of Eq. \ref{odd}.
Since $(3\beta+\gamma-2\delta)(\alpha-2\beta+3\delta)$ is positive 
when $n$ is odd, the ground state of superfluid phase is 
a polar state as the same as the case of even $n$. 


\begin{figure}
\includegraphics[width=\linewidth]{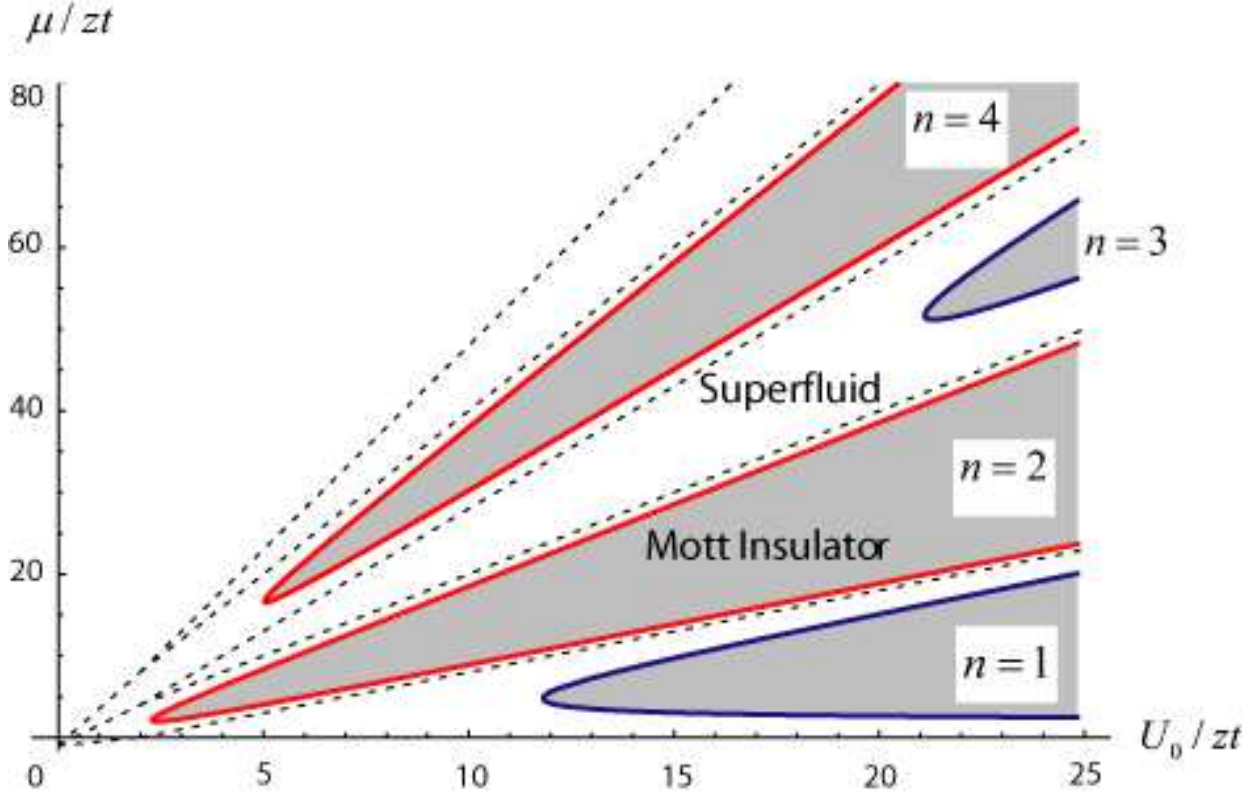}
\caption{\label{phase diagram}Phase diagram of bose-Hubbard 
model with spin 
degrees of freedom when $U_2/zt=1$. 
The dark region represents the Mott-insulating phase. 
The dashed lines indicate the phase boundaries at $t=0$.} 
\end{figure} 

Surprisingly, the Mott insulating phase with even $n$
is strongly stabilized against the superfluid phase 
comparing with odd-$n$ case. 
This is intuitively understood as follows: 
in the case of even number of atoms per site,  
all atoms are able to form singlet pairs on each site, 
while in the case of odd number atoms, 
one atom remains to be made pairing. 
In the former case,  
the boson pairs are strongly localized on a site
since the formation of singlet pairs prevent 
the bosons from hopping to the nearest-neighbor sites. 
Since the hopping is essential for 
the superfluid transition, the Mott insulating phase
is stabilized in the former case.  
On the contrary, 
remaining one atom is free to hop to 
the nearest-neighbor sites more freely in the latter case, 
thus the superfluid transition occurs easily.  
In addition, Eq. \ref{U0C} 
shows that $U_0^c$ decreases linearly with $U_2$, 
which is consistent with the above consideration 
since the formation of singlet pairs is energetically 
more favorable when $U_2$ is larger. 

This ``even-odd conjecture'' reminds us 
one-dimensional antiferromagnetic Heisenberg models 
(Haldane's conjecture) \cite{Haldane} 
and electronic ladder systems \cite{ladder} 
such as Hubbard or $t-J$ ladders. 
Both systems show similar properties; 
the Haldane (ladder) systems have a spin excitation gap
and an exponential decay of the spin correlation function
with an integer spin (an even number of legs), 
while gapless and power-low decay 
with a half-integer spin (an odd number of legs). 
These properties are able to be essentially explained in terms of 
tightly bound spin singlets as the present study. 
However, we have applied this even-odd conjecture to 
the Bose systems or S-I transitions for the first time. 

Finally we note that 
if we assume a ferromagnetic inter-atom interaction ($U_2<0$), 
there seems to be no strong even-odd  
dependence of the phase boundaries since the Mott insulator phase 
is the highest spin state and does not include singlet pairs. 
The detailed comparison between the ferromagnetic case 
and the present study 
remains as a future problem. 
Other possibilities such as fragmented condensates or 
two-particle pairings \cite{Kagan}
should also be studied in this system. 

To summarize, we have investigated the S-I 
transition of spin-1 bosons in an optical lattice.
The zero-temperature phase diagram has been obtained 
by a mean-field theory. 
We have determined the order parameter of superfluid phase 
and showed that superfluid phase is a polar state.
More interestingly, we have obtained a new kind of 
even-odd conjecture that the Mott insulating 
phase is strongly stabilized 
against the superfluid phase 
only when the number of atoms per site is even. 

We would like to thank M. Wadati, T. Nikuni, N. Hatakenaka, 
M. Nishida, N. Yokoshi and K. Kamide for fruitful discussions.  
This work was partly supported by COE 
Program ``Molecular Nano-Engineering'' from the Ministry of Education, 
Science and Culture, Japan. 
S.T. is supported by the Japan Society for Promotion of Science.


\end{document}